\documentclass[epj]{svjour}
\usepackage{latexsym}
\usepackage{graphicx}
\usepackage{color}
\usepackage{amssymb}
\usepackage{amsmath}
\usepackage[colorlinks=true,linkcolor=blue,citecolor=red]{hyperref}
\usepackage{float}
\usepackage{url}
\usepackage{subfig}
\usepackage{bm}
\usepackage{widetext}
\usepackage{multirow}
\usepackage{booktabs}
\usepackage{siunitx}

\def\be{\begin{equation}}
\def\ee{\end{equation}}
\def\ba{\begin{eqnarray}}
\def\ea{\end{eqnarray}}

\begin{document}

\title{Transverse Energy per Charged Particle in Heavy-Ion Collisions: Role of Collective Flow}

\author{Swatantra Kumar Tiwari\inst{1}, Raghunath Sahoo\inst{1,}\thanks{\emph email: Raghunath.Sahoo@cern.ch (corresponding author)}
}                     
%
%
\institute{Discipline of Physics, School of Basic Sciences, Indian Institute of Technology Indore, Simrol, Khandwa Road, Indore- 453552, India 
}
\date{\today}

\abstract{The ratio of (pseudo)rapidity density of transverse energy and the (pseudo)rapidity density of charged particles, which is a measure of the mean transverse energy per particle, is an important observable in high energy heavy-ion collisions. This ratio reveals information about the mechanism of particle production and the freeze-out criteria. Its collision energy and centrality dependence is almost similar to the chemical freeze-out temperature till top Relativistic Heavy-Ion Collider (RHIC) energy. The Large Hadron Collider (LHC) measurement at $\sqrt{s_{NN}}$ = 2.76 TeV brings up new challenges towards understanding the phenomena like gluon saturation and role of collective flow etc. being prevalent at high energies, which could contribute to the above observable. Statistical Hadron Gas Model (SHGM) with a static fireball approximation has been successful in describing both the centrality and energy dependence until top RHIC energies. However, the SHGM predictions for higher energies lie well below the LHC data. In order to understand this, we have incorporated collective flow in an excluded-volume SHGM (EV-SHGM). Our studies suggest that the collective flow plays an important role in describing $E_{T}/N_{ch}$ and it could be one of the possible parameters to explain the rise observed in $E_{T}/N_{ch}$ from RHIC to LHC energies. Predictions are made for $E_{T}/N_{ch}$, participant pair normalized transverse energy per unit rapidity and the Bjorken energy density for Pb+Pb collisions at $\sqrt{s_{NN}}$ = 5.02 TeV at the Large Hadron Collider.}
 

\authorrunning{S.~K.~Tiwari and R.~Sahoo}
\titlerunning{Transverse Energy per Charged Particle in Heavy-Ion Collisions}
\maketitle 
\section{Introduction}
Heavy-ion collision experiments at ultra-relativistic energies aim to produce a partonic phase of matter and study the Quantum Chromodynamics (QCD) deconfinement transition. Also, these experiments explore/scan the QCD phase diagram for a possible location of the critical point (CP), by controlling the temperature and baryochemical potential by changing the collision species/ centrality and collision energy. In these efforts, the future facilities like CBM experiment at FAIR energies ($E_{lab}$ : 10 AGeV-40 AGeV), the RHIC Beam Energy Scan programs, the LHC, and beyond (FCC, ILC) would play a pivotal role in exploring the QCD phase boundary, establishing the nature of the QCD phase transition and the location of CP.  The RHIC at the Brookhaven National Laboratory, USA has successfully discovered a strongly interacting partonic matter, which behaves like a liquid with the lowest $\eta/s$ ratio \cite{Arsene:2004fa,Back:2004je,Adams:2005dq,Adcox:2004mh}, which is comparable with the ADS/CFT calculations \cite{Policastro:2001yc}. The collision of small systems ($p+p$) at the LHC seems to show collectivity \cite{Khachatryan:2016txc}, which is a possible signature of Quark-Gluon Plasma (QGP) and was initially expected to be formed only in top central heavy-ion collisions. Along with the energy loss patterns of heavy-quarks in the medium formed at the LHC, these are few very important aspects of the new states of matter formed at ultra-relativistic energies.
In addition, the matter formed at LHC energies has been seen to show properties very different from that is observed at RHIC in many aspects, {\it i.e.} suppression of $J/\psi$ \cite{Abelev:2012rv} and other quarkonia, $\frac{dE_T}{d\eta(y)}/\frac{dN_{ch}}{d\eta(y)} \equiv E_{T}/N_{ch}$ showing behaviour not expected by equilibrium Statistical Hadron Gas Model (SHGM) with a static fireball approximation \cite{Cleymans:2007uk,Mishra:2013dwa,Sahoo:2014aca}, the collision energy dependence of $dN_{ch}/d\eta$ deviating from a logarithmic behaviour etc. There is a need to study the global observables like $E_{T}/N_{ch}$ in more details including the collectivity in the system to understand its energy and centrality dependence and to put some light on possible effects from other sources like the gluon saturation. 

The rapidity (pseudorapidity) density of transverse energy, $dE_T/$ $dy(\eta)$ is a measure of the energy distribution and explosiveness of the collision. This is an important observable, as it is the energy of the produced particles in the transverse phase space, which was completely empty before the collision. The energy of the incoming nuclei in the longitudinal direction is converted to the energy of the produced particles. The mid-rapidity measurement of $dE_T/dy$ is related to the initial energy density of the system. In longitudinal boost invariant Bjorken hydrodynamics \cite{Bjorken:1982qr}, this helps in making a direct comparison with the lattice QCD prediction of energy density for a deconfinement transition and thereby giving a first hint of a possible partonic medium. The study of the centrality and collision energy dependence of $dE_T/dy(\eta)$ sheds light on possible freeze-out criteria in heavy-ion collisions \cite{Cleymans:2007uk}.

In this paper, we study the variation of $E_{T}/N_{ch}$ with respect to centrality and centre-of-mass energy ($\sqrt{s_{NN}}$) over a broad energy range from 2.7 GeV to 5.02 TeV using EV-SHGM with collective flow. Experimentally, $E_{T}/N_{ch}$  increases rapidly at lower energies and then it saturates around SPS energies up to top RHIC energy. Till lower SPS energies, the increase in collision energy increases the mean energy or transverse mass ($m_T$) of particles. From SPS to RHIC energies, the additional energy pumped into the system in terms of the increase in $\sqrt{s_{NN}}$, goes towards new particle production \cite{Cleymans:2007uk,Mishra:2013dwa}. Recently, the experimental data at LHC energy of 2.76 TeV show a sharp rise in this spectrum due to further increase in mean energy or $m_T$ of the particles \cite{Mishra:2013dwa}, and possible collective effects. This behaviour of $E_{T}/N_{ch}$ as a function of collision energy does not follow a static fireball expectations \cite{Cleymans:2007uk}. In order to understand this at LHC energies, we use our recently proposed model, where  we incorporate the attractive interactions by including the resonances up to a mass of 2 GeV and repulsive interactions by assigning the geometrical hard-core size to baryons. Mesons are treated as pointlike particles in the model. We also incorporate the collective flow in the model to explain the experimental data at various energies particularly at LHC. In ref. \cite{Prorok:2004af}, the statistical model is also used to study the transverse energy per charged particle at mid-rapidity with longitudinal and transverse flows for a wide range of energies from AGS to RHIC. In our case, we extend our analysis up to LHC energies, where the role of collective flow is more pronounced than at lower energies. In addition, we study the centrality dependence of $E_{T}/N_{ch}$ at $\sqrt{s_{NN}}$ = 200 GeV and 2.76 TeV, which is related to the chemical freeze-out of the system. We study the energy dependence of the associated observables like the participant pair normalized $\displaystyle dE_T/dy(\eta)$, the Bjorken energy density ($\epsilon_{Bj}$) in order to study the created system at different energies and the possible different behaviour at LHC energies, which could serve the purpose of ruling out and/or establishing different production mechanisms. In our calculation, we assume that the chemical and thermal freeze-outs occur simultaneously which infers the absence of the possible elastic scattering after chemical freeze-out~\cite{Prorok:2004wi,Prorok:2006ve}. 
 
The paper is organized as follows: in section~\ref{formulation}, we give the formulation of the SHGM with an excluded volume correction and the method of inclusion of collective flow. In section~\ref{results}, we present the results and discussions. In section~\ref{summary}, we give the summary with outlook and open problems.
 
\section{Formulation of The Model}
\label{formulation}
The formula for the number density of the $i$-th baryon in the excluded-volume model using the Maxwell-Boltzmann's statistics is written as \cite{Mishra:2008tc}:
\begin{equation}
n_i^{ex} = (1-R)I_i\lambda_i-I_i\lambda_i^2\frac{\partial{R}}{\partial{\lambda_i}},
\label{eq1}
\end{equation}
where $R=\sum_in_i^{ex}V_i^0$ is the fractional occupied volume by the baryons \cite{Tiwari:2013wga}. $\displaystyle V_i^0= (4\pi\;r'^3)/3$ and $\lambda_i$ are the eigen-volume and fugacity of the $i$-th baryon having a hard-core radius $r'$, respectively. Here we take $r'$ = 0.8\;$fm$, which is a free parameter in the discussed model. $I_i$ is the momentum integral for baryons in the Boltzmann's statistics. Eq.~\ref{eq1} can be reduced in the following form \cite{Tiwari:2013pva,Tiwari:2013}:
\ba \label{eq2}
\frac{dN_i}{dy\;m_T\;dm_T\;d\phi_p}=\frac{g_iV\lambda_i}{(2\pi)^3}\Big[\Big((1-R)-\lambda_i\frac{\partial{R}}{\partial{\lambda_i}}\Big) \nonumber \\                                                   \times \frac{E_i}{\displaystyle \Big[\exp\left(\frac{E_i}{T}\right)\Big]}\Big].
\ea
Here $y$ is the rapidity variable and $m_T=\sqrt{{m}^2+{p_T}^2}$ is the transverse mass. $E_i$ is the energy of the $i$-th baryon, $V$ is the total volume of the fireball formed at chemical freeze-out and $N_i$ is the number of the $i$-th baryon. We assume that the volume of the fireball, $V$ is the same for all types of hadrons at the time of the homogeneous emissions.

By using $E_i=m_T{\cosh}y$, Eq.~\ref{eq2} can be written as \cite{Tiwari:2013}:
\ba \label{eq3}
\frac{dN_i}{dy\;m_T\;dm_T\;d\phi_p}=\frac{g_iV\lambda_i}{({2\pi})^3}\;\Big[\Big((1-R)-\lambda_i\frac{\partial{R}}{\partial{\lambda_i}}\Big) \nonumber \\                                           \times \frac{m_T\;{\cosh}y}{\displaystyle\Big[\exp\left(\frac{m_T\;{\cosh}y}{T}\right)\Big]}\Big].
\ea
\subsection{Transverse Energy of Hadrons in a Thermal Model}
The transverse energy, $E_T$ in an event is defined as: 
\begin{equation}
E_T = \sum_i E_i \sin \theta_i,
\label{eq4}
\end{equation} 
with $\theta_i$ as the polar angle made by the $i$-th particle in an event with the detector. The sum is taken over all the particles emitted into a fixed solid angle within the detector acceptance. Taking into account the calorimetry measurement of $E_T$, one redefines the energy of the individual particles as \cite{Adam:2016thv,Adams:2004cb,Adler:2004zn},

\begin{equation}\label{eq5}
       E_{\mathrm i} = \left\{
    \begin{array}{ll}
                       E_{\mathrm {total}}-m & \mbox{for baryons} \\
                       E_{\mathrm {total}} + m & \mbox{for anti-baryons} \\
                       E_{\mathrm {total}} & \mbox{for all other particles.}
    \end{array}
    \right.
    \end{equation}
Considering the above experimental formulae, we proceed with the formulation of the transverse energy in EV-SHGM. Using Eq.~\ref{eq3}, we write the energy of the $i$-th baryon per unit rapidity at mid-rapidity ($y$ = 0) as: 
\ba \label{eq6}
\Big(\frac{dE_i}{dy}\Big)_{y=0}=\frac{g_iV\lambda_i}{(2{\pi})^2}\;\Big[\Big((1-R)-\lambda_i\frac{\partial{R}}{\partial{\lambda_i}}\Big)\Big] \nonumber \\ 
\times \int \frac{m_T^{3}\;dm_{T}}{\displaystyle\Big[\exp\left(\frac{m_T}{T}\right)\Big]}.
\ea

Similarly, the energy of the $m$-th meson per unit rapidity at $y$ = 0 is calculated as:
\begin{equation}
\Big(\frac{dE_m}{dy}\Big)_{y=0}=\frac{g_mV\lambda_m}{(2{\pi})^2}\;\int \frac{m_T^{3}\;dm_{T}}{\displaystyle\Big[\exp\left(\frac{m_T}{T}\right)\Big]}.
\label{eq7}
\end{equation}
Here, $g_m$ and $\lambda_m$ are the degeneracy factor and fugacity of the $m$-th meson. The above equations give the energy of the particles arising from a stationary thermal source.
 
\subsection{Transverse Energy of Hadrons in a Thermal Model with Flow}
The invariant yield with the inclusions of collective flow in EV-SHGM can be written as~\cite{Cooper:1974mv}:

\begin{eqnarray}
E_i\;\frac{d^3N_i}{dp^3}=\frac{g_i\lambda_i}{(2\pi)^3}\;\Big[(1-R)-\lambda_i\frac{\partial{R}}{\partial{\lambda_i}}\Big] \nonumber \\
                                        \times\int exp\Big(\frac{-p^{\mu}u_{\mu}}{T}\Big)\;p^\lambda\;d\sigma_\lambda.
                                        \label{eq8}
\end{eqnarray}
The freeze-out hypersurface $d\sigma_{\lambda}$ in Eq. \ref{eq8} is parametrized in cylindrical coordinates $(r,\phi,\eta)$. In the derivation of Eq. \ref{eq8}, it is assumed that an isotropic thermal distribution of hadrons is boosted by the local fluid velocity, $u_{\mu}$. Here, four momentum ($p^\mu$) and $u_{\mu}$ are defined as:
\begin{eqnarray}
p^{\mu} = (m_T{\cosh}y, p_T{\cos}\phi, p_T{\sin}\phi, m_T{\sinh}y),
\label{eq9}
\end{eqnarray}
 and
\begin{eqnarray}
u_{\mu}(\rho,\eta) = {\cosh}\rho({\cosh}\eta, {\tanh}\rho, 0, {\sinh}\eta).
\label{eq10}
\end{eqnarray}
Now, Eq. \ref{eq8} becomes,
 
\begin{widetext}
\ba \label{eq11}
\frac{dN_i}{m_T\;dm_T\;dy\;d\phi_p}=\frac{g_i\lambda_i\;m_{T}}{(2\pi)^3}\;\Big[(1-R)-\lambda_i\frac{\partial{R}}{\partial{\lambda_i}}\Big]\;\int \exp\Big(-\frac{m_T {\cosh}(y-\eta)\;{\cosh}\rho-p_T{\sinh}\rho\;{\cos}\phi}{T}\Big)r\; dr\;d\phi\;d{\zeta}.
\ea
\end{widetext}
Now, we make an assumption that the longitudinal flow is boost invariant. In this case, the longitudinal flow rapidity, $\eta$ is equal to the energy-momentum rapidity, y $ i. e.$, $\eta$ = y. Thus, Eq. \ref{eq11} is written as~\cite{Yin:2017qhg}:

\begin{widetext}
\ba \label{eq12}
\frac{dN_i}{m_T\;dm_T\;dy\;d\phi_p}=\frac{g_i\lambda_i\tau_0\;m_{T}}{(2\pi)^3}\;\Big[(1-R)-\lambda_i\frac{\partial{R}}{\partial{\lambda_i}}\Big] \; \int \exp\Big(-\frac{m_T\;{\cosh}\rho-p_T{\sinh}\rho\;{\cos}\phi}{T}\Big)     
                     r\; dr\;d\phi,
\ea
\end{widetext}
where $\tau_0$ is the proper freeze-out time. In the case of an instant thermal freeze-out, Eq.~\ref{eq12} is reduced in a simpler form~\cite{qgp}, 

\ba \label{eq13}
\frac{dN_i}{m_T\;dm_T\;dy}=\frac{g_i\lambda_iV\;m_{T}}{4\pi^3}\;\Big[(1-R)-\lambda_i\frac{\partial{R}}{\partial{\lambda_i}}\Big]\nonumber \\    
                                              \times \exp\Big(-\frac{m_T\;{\cosh}\rho}{T}\Big)\; I_0\Big(\frac{p_T{\sinh}\rho}{T}\Big).
\ea

Here, $\rho$ is the parameter given by $\rho={\tanh}^{-1}\beta_r$, $\beta_r$ is the radial flow. $V$ is the volume of the cylindrical matter. We assume that $T$ and $\rho$ are $r$-independent. $I_0$ is the modified Bessel function given as,
\ba \label{eq14}
\centering
I_0\Big(\frac{p_T\;{\sinh}\rho}{T}\Big)=\frac{1}{2\pi}\int_0^{2\pi} exp\Big(\frac{p_T\;{\sinh}\rho\;{\cos}\phi}{T}\Big)d\phi.
\ea
\\\
After incorporating the collective flow in thermal model, we get the expressions for energy of baryons per unit rapidity at $y$ = 0 as follows:

\begin{widetext}
\ba \label{eq15}
\Big(\frac{dE_i}{dy}\Big)_{y=0}=\frac{g_iV\lambda_i}{4\pi^3}\;\Big[(1-R)-\lambda_i\frac{\partial{R}}{\partial{\lambda_i}}\Big]\;\int \exp\Big(- \frac{m_T{\cosh}\rho}{T}\Big)\;I_0\Big(\frac{p_T{\sinh}\rho}{T}\Big)\;m_{T}^{3}\;dm_{T}.
\ea
\end{widetext}
In a similar fashion, we can calculate the energy per unit rapidity at mid-rapidity of the $m^{th}$ meson as:

\ba \label{eq16}
\Big(\frac{dE_m}{dy}\Big)_{y=0}=\frac{g_mV\lambda_m}{(4\pi^3)}\;\int \exp\Big(-\frac{m_T\;{\cosh}\rho}{T}\Big)\nonumber\\
                            \times I_0\Big(\frac{p_T{\sinh}\rho}{T}\Big)\;m_{T}^{3}\;dm_{T}.
\ea

Here, $E_{m}$, $g_m$, and $\lambda_m$ represent the energy, degeneracy factor and fugacity of the $m^{th}$ meson. Now, Eq.~\ref{eq4} can be reduced in the following form:

\begin{equation}
\langle E_{T} \rangle=\langle \sum_i E_i \sin \theta_i \rangle.
\label{eq17}
\end{equation} 
The average of $\sin\theta$ can be calculated as follows :
\begin{eqnarray} \label{eq18}
\langle \sin\theta \rangle=\frac{1}{4\pi}\int\; \sin\theta d\Omega 
&=&\frac{1}{4\pi}\int\; \sin^2\theta\; d\theta d\phi,
\end{eqnarray}
where $d\Omega$ ($=\sin\theta d\theta d\phi$) is the solid angle. Now, integrating the above equation for the pseudo-rapidity interval i. e. $|\eta|$ $<$ 0.88, which almost lies in the mid-rapidity region,
\begin{eqnarray} \label{eq19}
\langle\sin\theta\rangle=\frac{1}{4\pi}\int_{\pi/4}^{3\pi/4}\; \sin^2\theta\; d\theta \int_0^{2\pi} d\phi 
&=&\Big(\frac{\pi}{8}+\frac{1}{4}\Big).
\end{eqnarray}
We can write the expression of the transverse energy of hadrons using the above equation as follows:
\begin{equation} \label{eq20}
\langle E_{T} \rangle=\Big(\frac{\pi}{8}+\frac{1}{4}\Big)\Big[\langle E \rangle - m_{N}\langle N_{B} - N_{\bar{B}} \rangle\Big].
\end{equation} 
$\langle E \rangle$ is the total energy of hadrons at mid-rapidity. $N_{B} - N_{\bar{B}}$ is the net-baryon at $y$ = 0 which can be calculated by using Eq.~\ref{eq3}. After obtaining the transverse energy, we calculate the Bjorken energy density using the following formula~\cite{Bjorken:1982qr}:
\begin{equation} 
\epsilon_{Bj}=\frac{dE_{T}}{dy}\frac{1}{\tau\;\pi\;R^{2}},
\label{eq21}
\end{equation} 
where $\tau$ is the formation time and $\pi\;R^{2}$ is the transverse overlap area of the colliding nuclei. There are various ways to quantify the overlap area. Here, R is the radius of the colliding nuclei given by $R=R_{0}\;A^{1/3}$. Replacing $A$ by $N_{part}/2$, where $N_{part}$ is the number of nucleon participants~\cite{Kharzeev:2000ph}, $\epsilon_{Bj}$ becomes 
\begin{equation}
\epsilon_{Bj}=\frac{dE_{T}}{dy}\frac{1}{\tau\;\pi\;R_{0}^{2}\;\Big(N_{part}/2\Big)^{2/3}}.
\label{eq22}
\end{equation}

\section{Results and Discussions}
\label{results}
In this section, we provide the results calculated using the model described above and present the discussions.
\subsection{Estimation of $N_{ch}$ in EV-SHGM}
\begin{figure}
\includegraphics[height=22em]{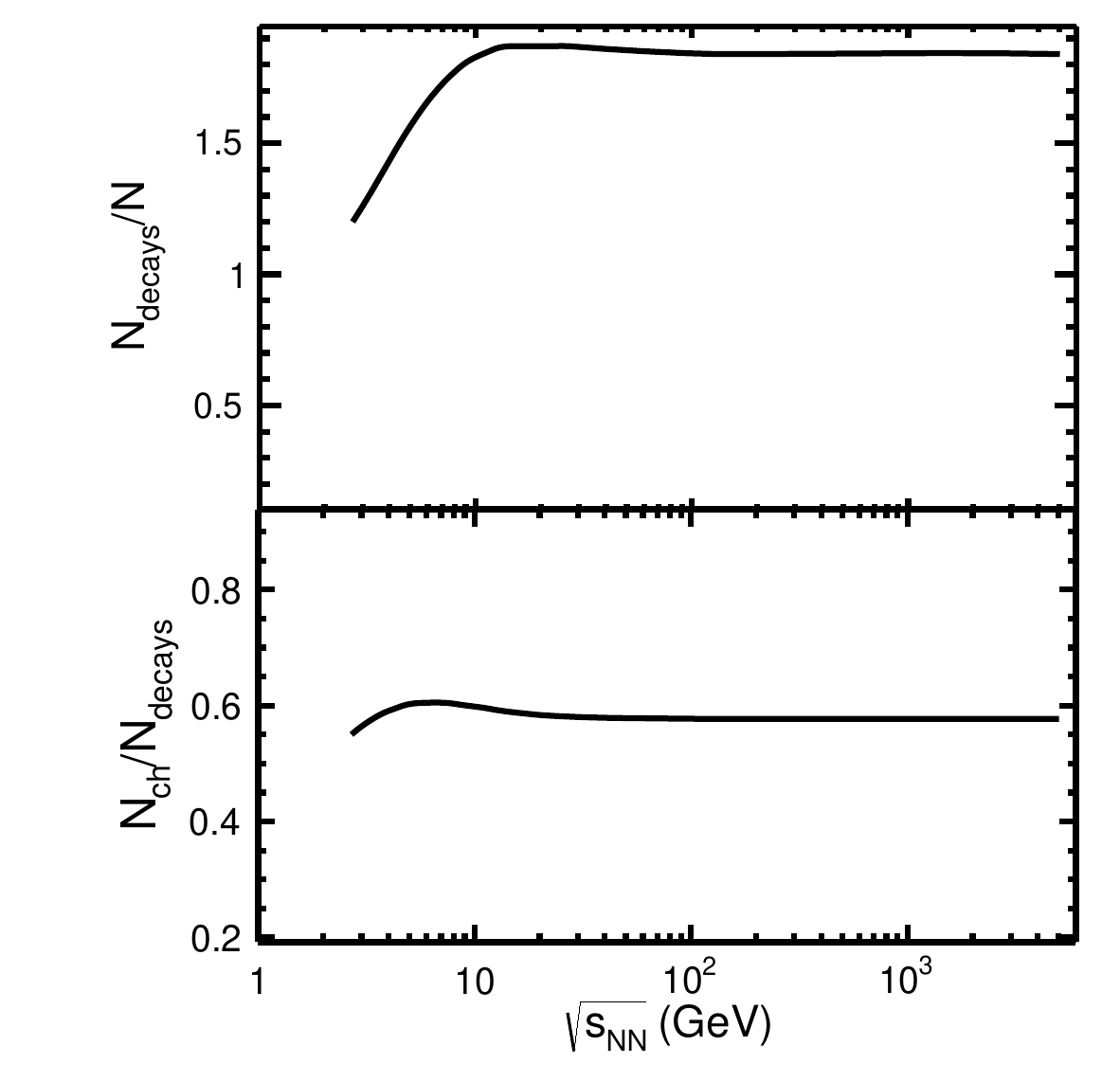}
\caption[]{The variation of $N_{decays}/N$ (upper panel) and $N_{ch}/N_{decays}$ (lower panel) at mid-rapidity ($y=0$) with $\sqrt{s_{NN}}$ at chemical freeze-out.}
\label{nch}
\end{figure}

In order to calculate the ratio $E_{T}/N_{ch}$, we estimate $N_{ch}$ at mid-rapidity in terms of the number of primarily produced particles, $N$. We follow the same procedure for the estimation of $N_{ch}$ as discussed in ref. \cite{Cleymans:2007uk}. We first calculate the ratio of the total number of hadrons in final state, $N_{decays}$ to the total number of primordial hadrons, $N$ at $y$ = 0 with respect to $\sqrt{s_{NN}}$ over a broad energy range from $\sqrt{s_{NN}}$ = 2.7 GeV to 5.02 TeV using EV-SHGM as shown in the upper panel of figure~\ref{nch}. Now, we study the ratio of the number of charged hadrons, $N_{ch}$ to $N_{decays}$ at mid-rapidity with $\sqrt{s_{NN}}$ from 2.7 GeV to 5.02 TeV, which is shown in the lower panel of figure~\ref{nch}. In order to calculate these observables in the framework of EV-SHGM, one needs the chemical freeze-out temperature ($T$), and baryon chemical potential ($\mu_{B}$) at each $\sqrt{s_{NN}}$ as mentioned in references \cite{Mishra:2008tc,Tiwari:2012}. For the centrality studies at RHIC and LHC energies, we estimate $T$ and $\mu_{B}$ by taking the best matching of the particle ratios between the experimental data and the calculations done in the framework of EV-SHGM for a given centrality class. These $T$ and $\mu_{B}$ are then used for the estimation of other observables discussed in the paper.
We find that the ratio $N_{decays}/N$ initially increases rapidly with $\sqrt{s_{NN}}$, because the production of resonances increases with energy and becomes saturated at SPS energies, where chemical freeze-out temperature becomes independent of collision energy. Similarly, the ratio $N_{ch}/N_{decays}$ also increases with $\sqrt{s_{NN}}$ and gets saturated at SPS energies. Although, these findings are the same as observed in Ref.~\cite{Cleymans:2007uk}, the difference occurs at lower energies where the excluded-volume correction is more effective.      

\subsection{Transverse Momentum Spectra and Extraction of Radial Flow}

\begin{table*}[tp]
 \centering
  \caption{The chemical freeze-out volume for $\pi^-$ at mid-rapidity for various center-of-mass energies.}
  \label{tab:table1}
  \begin{tabular}{c|c}
 \hline
$\sqrt {s_{NN}}$ (GeV)         &  $V$ ($\rm fm^3$) ($y$=0)   \\
  \hline
  $2.7$   	& 5532		\\
  $3.32$   	& 5446		\\
  $3.84$   	& 4703		\\
  $4.85$   	& 3350		\\	
$7.7$   	& 1712		\\
$11.5$ 	& 1172		 \\
 $19.6$     &  1024		\\
 $27$        &  1041	       \\
 $39$        &  1087		\\
 $62.4$     &  1402		\\
 $130$      & 1659		\\
 $200$      &  1940		\\
 $2760$      &  4355		\\
 $5020$      &  5476		\\
 \hline
\end{tabular}
\label{t1}
\end{table*}

\begin{figure}
\includegraphics[height=20em]{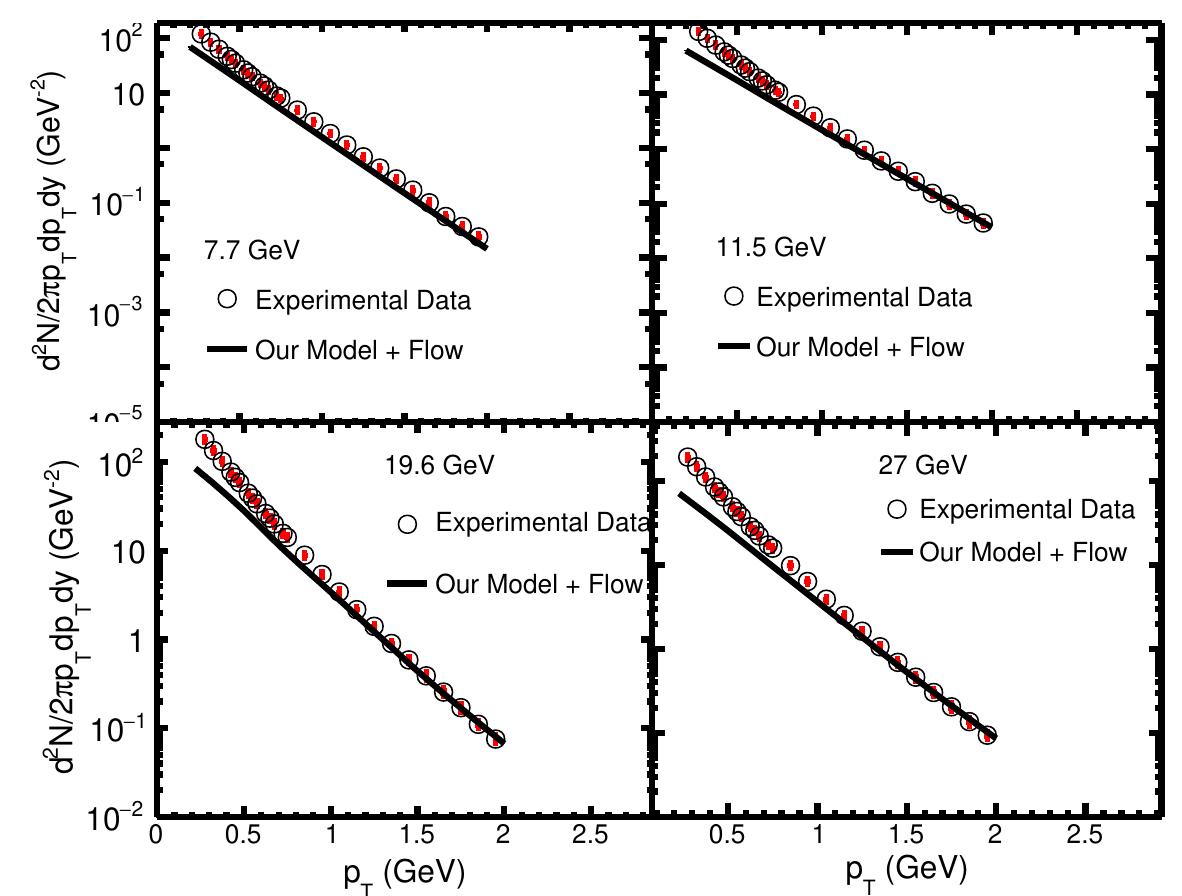}
\caption[]{The transverse momentum spectra of $\pi^{-}$ at $\sqrt{s_{NN}}$ = 7.7, 11.5, 19.6 and 27 GeV. Symbols are experimental data \cite{Adamczyk:2017iwn} while lines are model calculations.}
\label{pT}
\end{figure}

\begin{figure}
\includegraphics[height=20em]{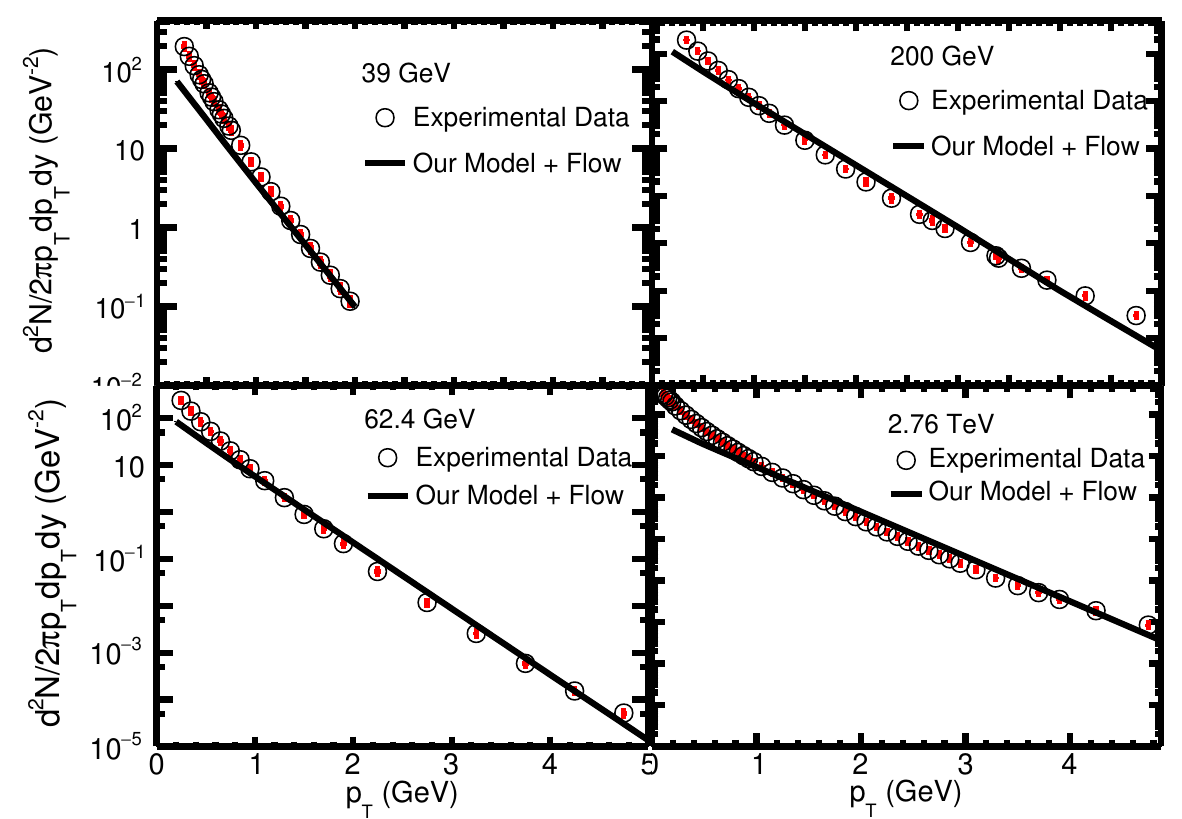}
\caption[]{The transverse momentum spectra of $\pi^{-}$ at $\sqrt{s_{NN}}$ = 39, 62.4, 200 GeV and ($\pi^{-}$+$\pi^{+}$) at $\sqrt{s_{NN}}$ = 2.76 TeV. Symbols are experimental data \cite{Adamczyk:2017iwn,Abelev:2007ra,Abelev:2006jr,Abelev:2014laa} while lines are model results.}
\label{pT1}
\end{figure}
We have estimated using the model the chemical freeze-out volume, $V$ of $\pi^-$ at various center-of-mass energies, which are tabulated in table~\ref{t1}. For this, we divide the experimental mid-rapidity density of $\pi^-$ at a particular $\sqrt{s_{NN}}$ to the corresponding value calculated in the model. In figures~\ref{pT} and~\ref{pT1}, we show the $p_{T}$ spectra of $\pi^{-}$ for the most central collisions at various $\sqrt{s_{NN}}$ from 7.7 GeV to 2.76 TeV. We use Eq.~\ref{eq13} for $\pi^-$ to calculate $p_{T}$-spectra, where $T$ and $\mu_B$ are taken from references.~\cite{Mishra:2008tc,Tiwari:2012}. We compare the results with the experimental data \cite{Adamczyk:2017iwn,Abelev:2007ra,Abelev:2006jr,Abelev:2014laa} up to $p_T$ = 5 GeV and find a good agreement between them. After comparison with the experimental data, we get the value of the radial flow velocity, $\beta_{r}$. In this paper, we do not take the contributions of resonance decays while calculating $p_T$ spectra, which may be the reason for a deviation observed at lower $p_{T}$~\cite{Chatterjee:2015fua}. This could be explored in a future work.

\begin{figure}
\includegraphics[height=25em]{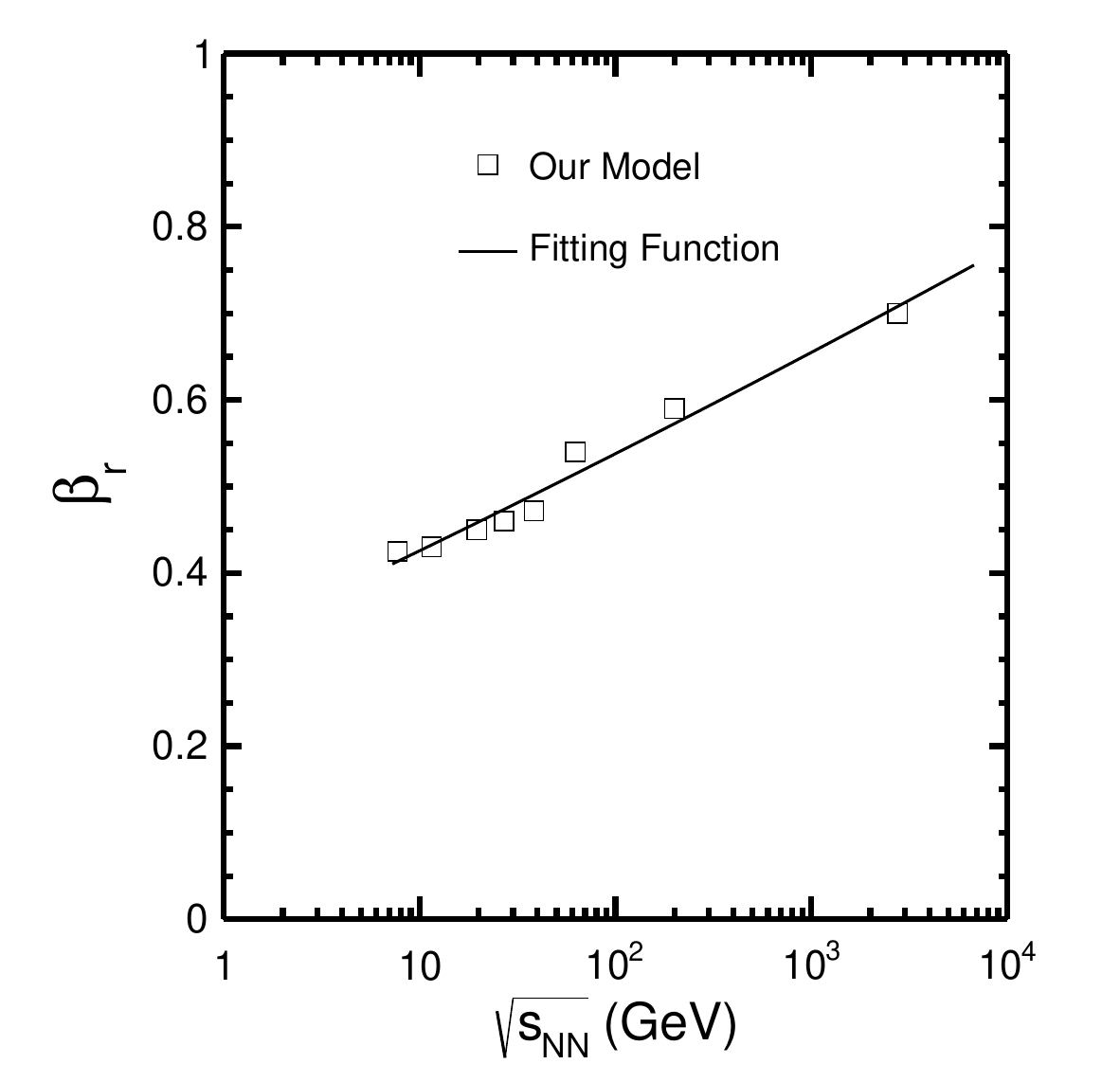}
\caption[]{The variations of the radial flow velocity ($\beta_r$) extracted in our model with respect to $\sqrt{s_{NN}}$. The solid line is a power-law fit as mentioned in the text.}
\label{flow}
\end{figure}
Figure~\ref{flow} shows the variations of radial flow velocity, $\beta_r$ extracted by fitting the $p_T$ spectra of $\pi^-$ in the framework of EV-SHGM with respect to $\sqrt{s_{NN}}$ from 7.7 GeV to 2.76 TeV. We notice that it increases monotonically with the collision energy with it's lowest value at 7.7 GeV to the highest value at LHC. This shows significant collectivity in high energy heavy-ion collisions. We fit a phenomenologically motivated power-law function, {\it i. e.} $\displaystyle a + b(\sqrt{s_{NN}})^{c}$ to the energy dependence of $\beta_r$. Here, a, b and c are the fit parameters. For the best fit we get, a = -2.206 $\pm$ 0.087, b = 2.5243 $\pm$ 0.0867, and c = 0.0181 $\pm$ 0.0014. The predicted value of $\beta_r$ at $\sqrt{s_{NN}}$ = 5.02 TeV is 0.73 for the top central Pb+Pb collisions. The extrapolated $\beta_r$ for $\sqrt{s_{NN}}$ = 2.7 GeV is 0.36. It should be noted here that, in the discussed energy domain although the radial flow shows a linear increase with collision energy, as it is expected, it has to start saturating at some higher energies in order to satisfy the limit of speed of light, $c$.

\subsection{Transverse Energy and Bjorken Energy Density}

\begin{figure}
\includegraphics[height=25em]{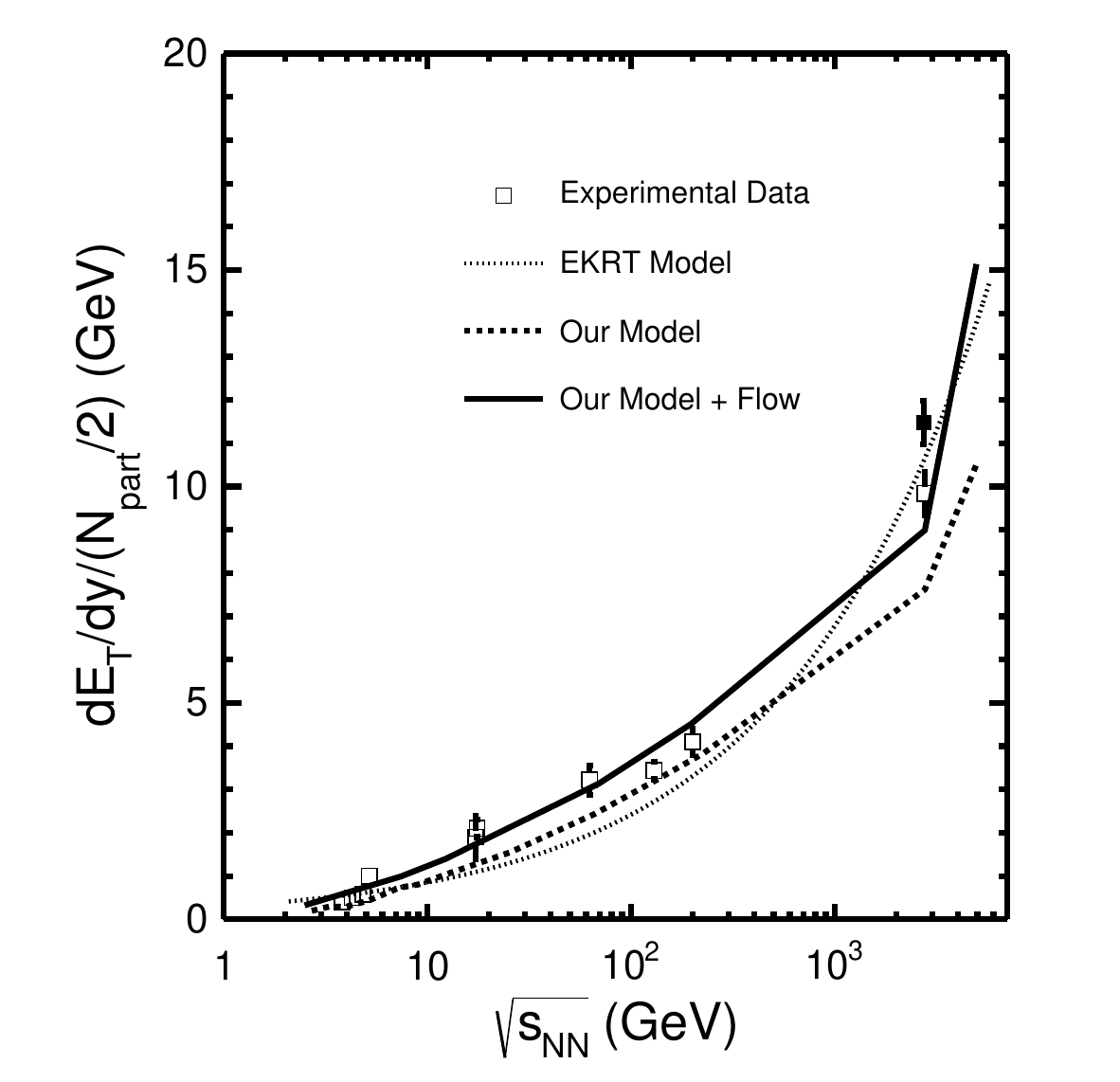}
\caption[]{The variation of $dE_T/dy$ per number of participant with respect to $\sqrt{s_{NN}}$ for the most central collisions. The solid line represents our model calculation with flow and the dashed line is the result obtained in our model without flow. The dotted line presents the results of EKRT model. Symbols are the experimental data points \cite{Sahoo:2014aca,Adam:2016thv,Adler:2004zn,Chatrchyan:2012mb}. The values are calculated at various discrete energies.}
\label{ET_NP}
\end{figure}

In figure~\ref{ET_NP}, we present the variations of $(dE_T/dy)/0.5N_{part}$ with respect to $\sqrt{s_{NN}}$ over a broad energy range from 2.7 GeV to 5.02 TeV. We calculate $((dE_T/dy)/0.5N_{part})$ using Eq.~\ref{eq20} and compare with the experimental data \cite{Sahoo:2014aca,Adam:2016thv,Adler:2004zn,Chatrchyan:2012mb}. The open symbol in the figure at $\sqrt{s_{NN}}$ = 2.76 TeV represents the ALICE data  \cite{Adam:2016thv} while solid symbol is the measurement by the CMS experiment \cite{Chatrchyan:2012mb}. Here, we take care of conversion of the $dE_T/d\eta$ to $dE_T/dy$ at LHC by using the Jacobian factor J($\eta$,y), which is 1.09 at this energy \cite{Chatrchyan:2012mb}. We notice that the model with flow explains the  ALICE data within experimental error but lies below to CMS data at LHC. We also show the results obtained in the EKRT model \cite{Eskola:1999fc} which is based on the calculation of perturbative QCD with gluon saturation mechanism and hydrodynamics. We observe that the EKRT model lies below the experimental data up to top RHIC energy but seems to explain the data at LHC energies.

 \begin{figure}
\includegraphics[height=25em]{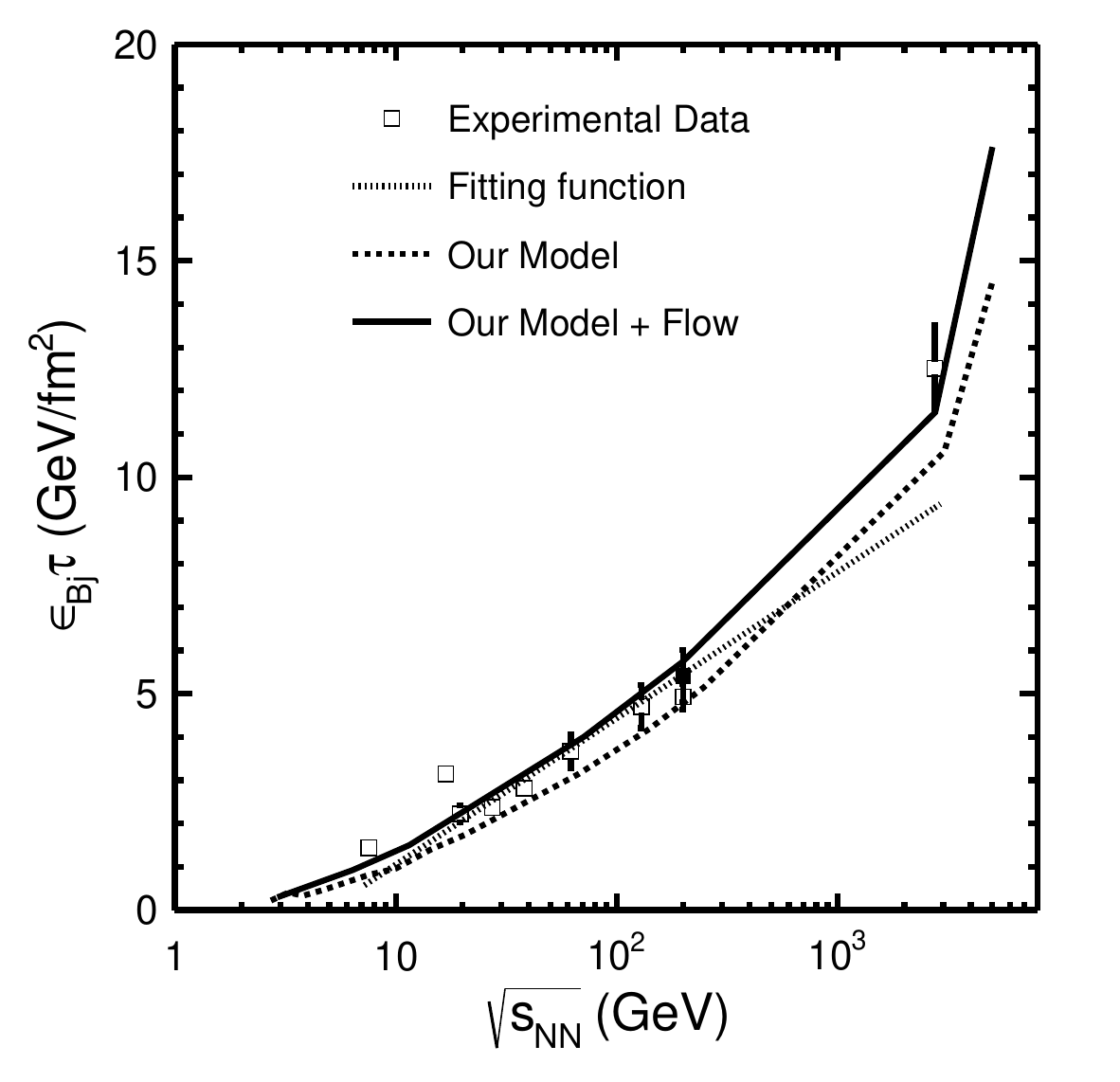}
\caption[]{The variation of $\epsilon_{Bj}\tau$ with respect to $\sqrt{s_{NN}}$ for the most central collisions. The solid line represents our model calculation with flow and the dashed line is the result obtained in the stationary thermal model. The dotted line represents the result of the logarithmic fitting function as described in the text. Symbols are the experimental data points \cite{Sahoo:2014aca,Adam:2016thv,Adler:2004zn}.}
\label{Ebj}
\end{figure}
Figure~\ref{Ebj} represents the variations of the product of Bjorken energy density ($\epsilon_{Bj}$) and formation time ($\tau$) with respect to $\sqrt{s_{NN}}$ from 2.7 GeV to 5.02 TeV. Furthermore, the model with flow explains the experimental data \cite{Sahoo:2014aca,Adam:2016thv,Adler:2004zn} satisfactorily. We also fit the experimental data using the logarithmic function $\displaystyle A + B\;ln(\sqrt{s_{NN}}/\sqrt{s_0})$, where A = - 2.32 $\pm$ 0.51 $GeV/fm^2$, B=1.46 $\pm$ 0.12 $GeV/fm^2$ are fit parameters and we take $\sqrt{s_0}$ = 1 GeV. We notice that this function fits the data only upto RHIC energies and fails at LHC energies which suggests that logarithmic behaviour is not valid at LHC energies in this case. This could be an indication of a different particle production mechanism playing a role at LHC energies, which needs further investigations.
\begin{figure}
\includegraphics[height=25em]{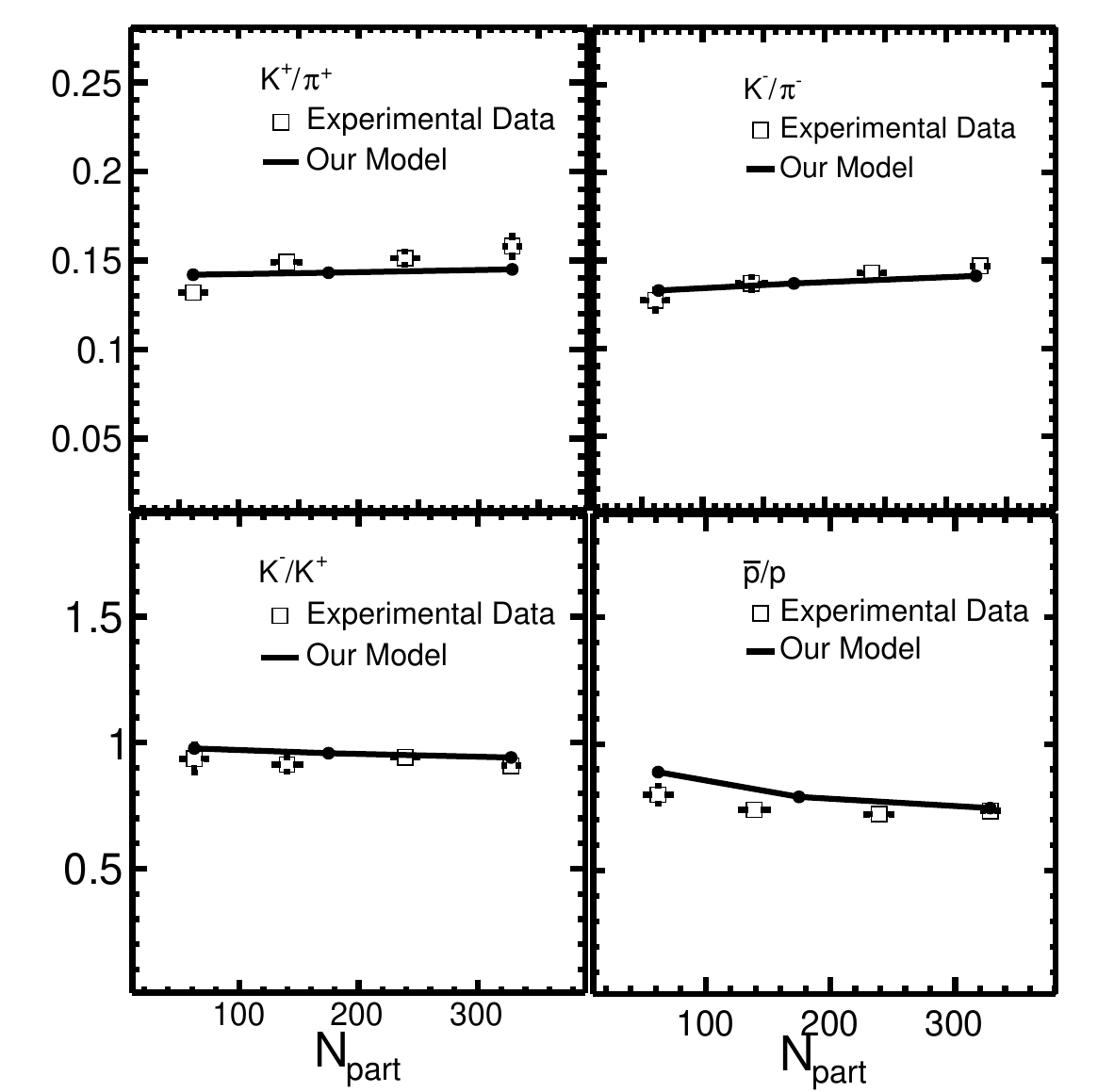}
\caption[]{The variations of particle ratios for Au-Au collisions with respect to number of participants at $\sqrt{s_{NN}}$ = 200 GeV. Symbols show the experimental data \cite{Arsene:2005mr} and lines are model calculations.}
\label{ratio}
\end{figure}

\subsection{Centrality Dependence of $E_{T}/N_{ch}$}

In order to study the variations of $E_{T}/N_{ch}$ with respect to centrality for various $\sqrt{s_{NN}}$, we need to estimate $T$ and $\mu_{B}$ for a given centrality class. To do this, we take various experimentally measured identified particle ratios at different centralities and match the corresponding particle ratios, estimated by using the proposed model with $T$ and $\mu_{B}$ as inputs. Here, we have taken the contributions from resonance decays while calculating the particle ratios. The best comparison gives the approximate value of $T$ and $\mu_{B}$ for a given centrality, which is represented by the number of participants ($N_{part}$). Then one uses these $T$ and $\mu_{B}$ values for the estimation of $E_{T}/N_{ch}$ in the framework of EV-SHGM. Figure~\ref{ratio} represents various hadron ratios such as $K^{+}/\pi^{+}$, $K^{-}/\pi^{-}$, $K^{-}/K^{+}$, and $\bar{p}/p$ \cite{Arsene:2005mr} with respect to the number of participants, $N_{part}$ for Au-Au collisions at $\sqrt{s_{NN}}$ = 200 GeV. We select three centrality bins with participant numbers 328$\pm$6 (most-central), 140$\pm$11 (mid-central), and 62$\pm$10 (peripheral) while calculating the particle ratios. We find that $T$ does not vary much with the centrality while $\mu_{B}$ decreases rapidly from most central to peripheral collisions. We find that our model explains the data very well over all the centralities. For the sake of convenience, we do not take the strangeness saturation factor ($\gamma_{s}$) into account in our analysis. In Ref. \cite{Cleymans:2004pp}, the detailed analysis of variations of particle ratios with centrality is done using $\gamma_{s}$. For $\sqrt{s_{NN}}$ = 2.76 TeV, we adopt a similar method as discussed above to estimate centrality dependent $T$ and $\mu_{B}$, which are further used for the estimation of $E_{T}/N_{ch}$ at the LHC. The extracted values of $T$ and $\mu_B$ at this energy are tabulated in table~\ref{t2}.

\begin{table*}[tp]
 \centering
  \caption{ Temperature and Baryon Chemical Potential extracted after fitting the particle ratios for various centrality at $\sqrt{s_{NN}}$ = 200 GeV and 2.76 TeV. Most-central, mid-central and peripheral are defined in the text for both energies.}
  \label{table2}
 \begin{tabular}{lSSSS}
    \toprule
    \multirow{2}{*}{Centrality} &
      \multicolumn{2}{c}{$\sqrt{s_{NN}}$ = 200 GeV} &
      \multicolumn{2}{c}{$\sqrt{s_{NN}}$ = 2.76 TeV} \\
      & {$T$ (MeV)} & {$\mu_B$ (MeV)} & {$T$ (MeV)} & {$\mu_B$ (MeV)} \\
      \midrule
    Most-central & 169 & 23.5 & 169 & 1.7\\
    Mid-central & 168.5 & 17 & 168.5 & 1.0  \\
    Peripheral & 168 & 5.5 & 168 & 0.5 \\
    \bottomrule
  \end{tabular}
\label{t2}
\end{table*}

 \begin{figure}
\includegraphics[height=25em]{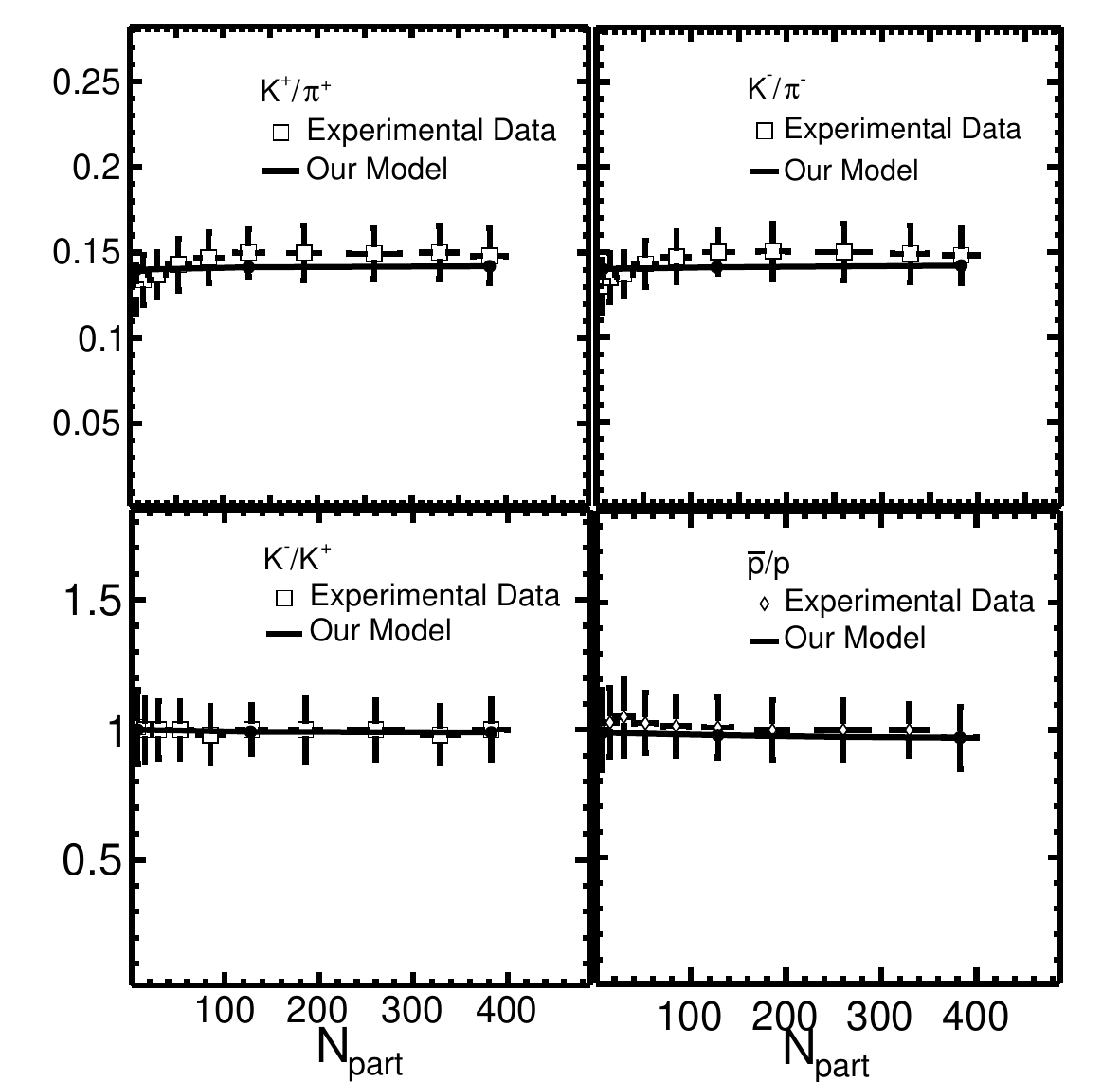}
\caption[]{The variations of hadron ratios for Pb-Pb collisions with respect to $N_{part}$ at $\sqrt{s_{NN}}$ = 2.76 TeV. Symbols show the experimental data \cite{Abelev:2013vea} and lines are our model results.}
\label{ratio_LHC}
\end{figure}
Figure~\ref{ratio_LHC} represents the centrality dependence of various hadrons ratios for Pb-Pb collisions at $\sqrt{s_{NN}}$ = 2.76 TeV. Again, we find that $T$ and $\mu_{B}$ do not vary much with centrality at this energy. While calculating ratios with respect to $N_{part}$, we select three centrality bins in our model with participant numbers 382$\pm$17 (most-central), 128$\pm$16 (mid-central), and 7$\pm$4 (peripheral). We do not include $\gamma_{s}$ in this analysis for the sake of simplicity. We compare our results with the experimental data \cite{Abelev:2013vea} and find that the model explains the data very well over all the centralities. Again, $T$ and $\mu_B$ extracted at this energy are tabulated in table~\ref{t2}

\begin{figure}
\includegraphics[height=25em]{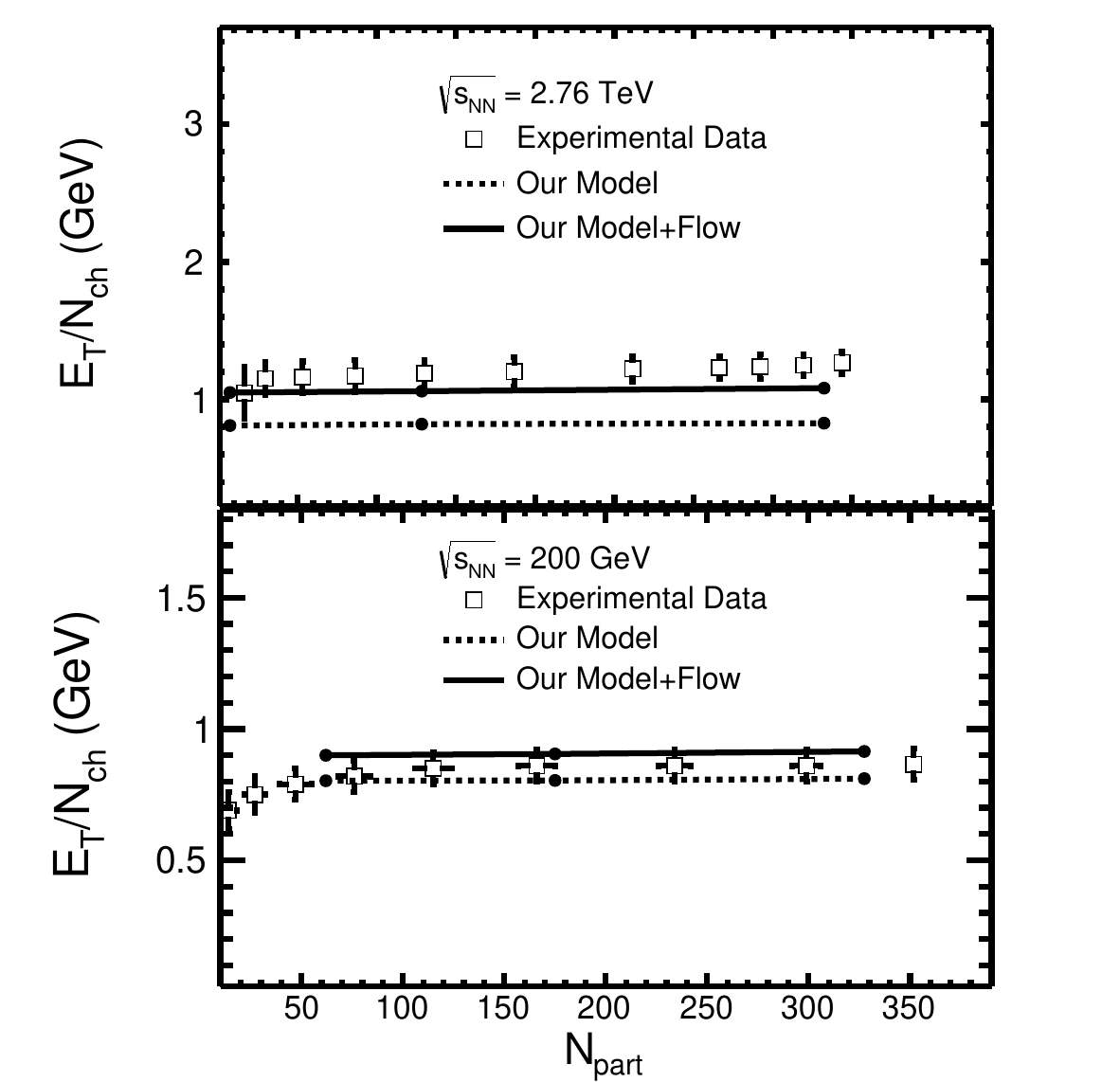}
\caption[]{Transverse energy per charged particle versus number of participants for Au-Au collisions at $\sqrt{s_{NN}}$ = 200 GeV (bottom) and for Pb-Pb collisions at $\sqrt{s_{NN}}$ = 2.76 TeV (top). Symbols are the experimental data \cite{Adam:2016thv,Adams:2004cb}. Solid lines are the results obtained in our model with flow while dotted lines are for the model without flow.}
\label{ET_centrality}
\end{figure}
In figure~\ref{ET_centrality}, we show the variations of $E_{T}/N_{ch}$ with $N_{part}$. In the upper panel, we show the results  for Pb-Pb collisions at $\sqrt{s_{NN}}$=2.76 TeV. We calculate $E_{T}/N_{ch}$ by using the values of $T$ and $\mu_{B}$ given in the table~\ref{t2}. We compare our results with the experimental data and find that EV-SHGM with flow describes the experimental data \cite{Adam:2016thv} within the experimental errors. In the lower panel, the results for Au-Au collisions at $\sqrt{s_{NN}}$ = 200 GeV are shown. Again, we take the values of $T$ and $\mu_{B}$ as given in the table~\ref{t2} while studying the $N_{part}$ dependence of $E_{T}/N_{ch}$ at this energy. We compare our results with the experimental data \cite{Adam:2016thv,Adams:2004cb} and find a very good agreement. In our calculations, we take the same centrality bins as used in the calculation of particle ratios at these energies. $E_{T}/N_{ch}$ is almost independent of centrality except at lower centrality bins. The present model explains the data well except at a lower $N_{part}$. In Ref.~\cite{Heinz:2007in}, it is argued that in a kinetic freeze-out scenario, the temperature is centrality dependent because during the kinetic freeze-out process there is a competition between local scattering and global expansion. Thus, the kinetic temperature is sensitive to the freeze-out process and hence becomes centrality dependent. In the case of chemical freeze-out, the temperature is observed to be centrality independent \cite{Adams:2003xp}. This is because during this process, the chemical reactions decrease abruptly leaving behind the chemically frozen state at the freeze-out and thus the chemical freeze-out temperature is insensitive to the collective dynamics but depends on thermodynamical variables. The observation of a centrality independence of $E_{T}/N_{ch}$ at RHIC and LHC thus indicates a chemical freeze-out scenario. This argument could be strengthened further in the subsequent section, when we make a direct comparison of $E_{T}/N_{ch}$ values with the universal freeze-out criteria.

\subsection{Energy Dependence of $E_{T}/N_{ch}$}
\begin{figure}
\includegraphics[height=25em]{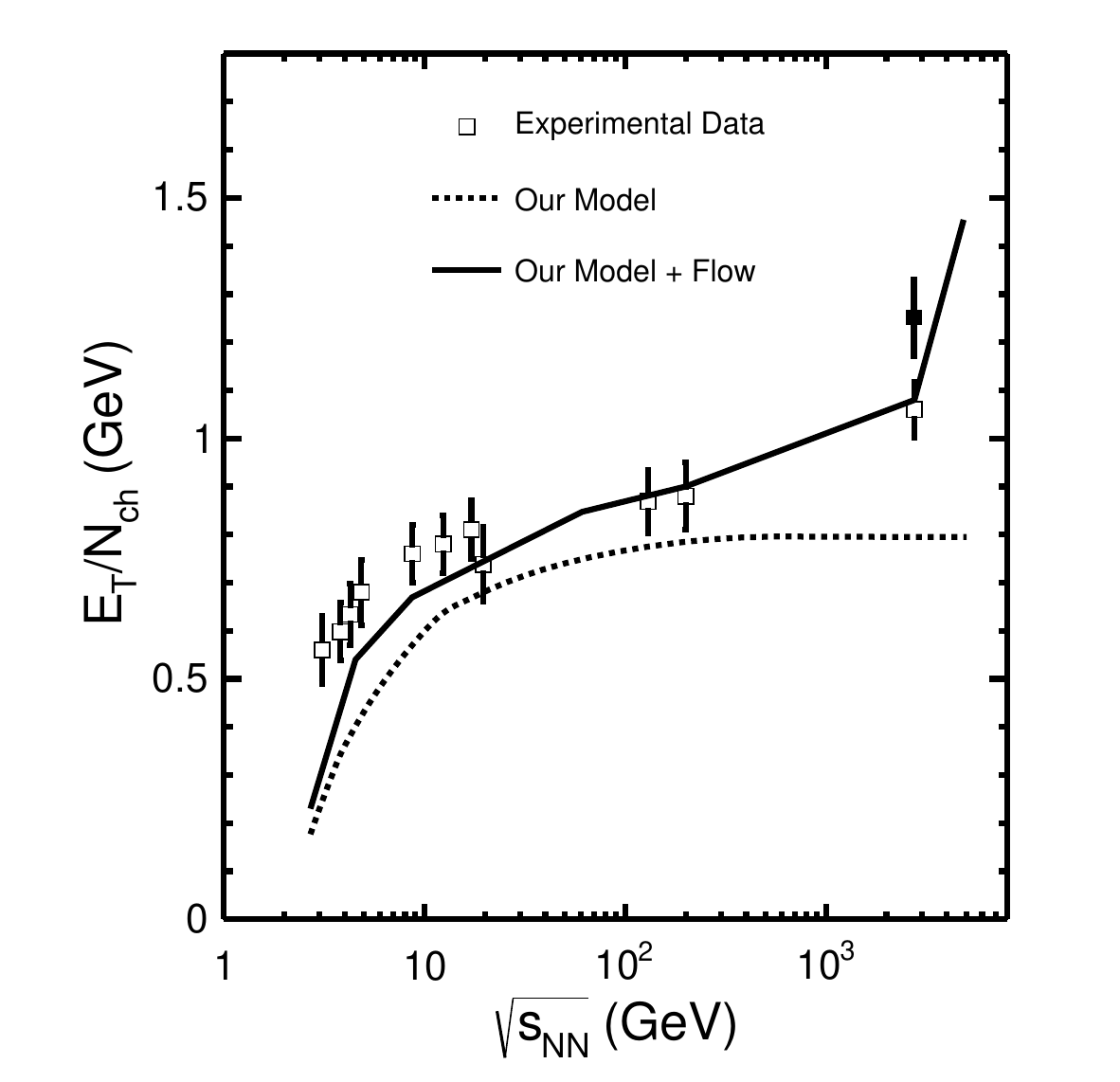}
\caption[]{Transverse energy per charged particle ($E_{T}/N_{ch}$) as a function of collision energy $\sqrt{s_{NN}}$. Experimental data \cite{Sahoo:2014aca,Adam:2016thv,Adler:2004zn,Chatrchyan:2012mb} are compared with the expectations from EV-SHGM with and without the effect of collective flow. The values are calculated at discrete energies.}
\label{et_snn}
\end{figure}
In figure~\ref{et_snn}, we demonstrate the ratio $E_{T}/N_{ch}$ for the most-central collisions with respect to $\sqrt{s_{NN}}$ starting from 2.7 GeV to 5.02 TeV. We confront EV-SHGM calculations with the experimental data \cite{Sahoo:2014aca,Adam:2016thv,Adler:2004zn,Chatrchyan:2012mb}. Here, the solid symbol at LHC is the CMS data \cite{Chatrchyan:2012mb} while open symbol is the ALICE data \cite{Adler:2004zn}. The thermal model without flow seems to explain the data at SPS and RHIC energies qualitatively but fails at LHC energies. These findings may hint for a possible effect of collective flow, which plays an important role at the LHC. We notice that our model with flow explains the ALICE data \cite{Adam:2016thv} within the experimental errors whereas the CMS data  \cite{Chatrchyan:2012mb} stays a little higher than EV-SHGM with flow. This hints for a more precise estimation of $E_{T}/N_{ch}$ at the LHC. We have shown the predictions for $E_{T}/N_{ch}$ in Pb+Pb collisions at $\sqrt{s_{NN}}$=5.02 TeV using the extrapolated value of $\beta_r$ as discussed above.

\subsection{$E_{T}/N_{ch}$ and Freeze-out}
\begin{figure}
\includegraphics[height=25em]{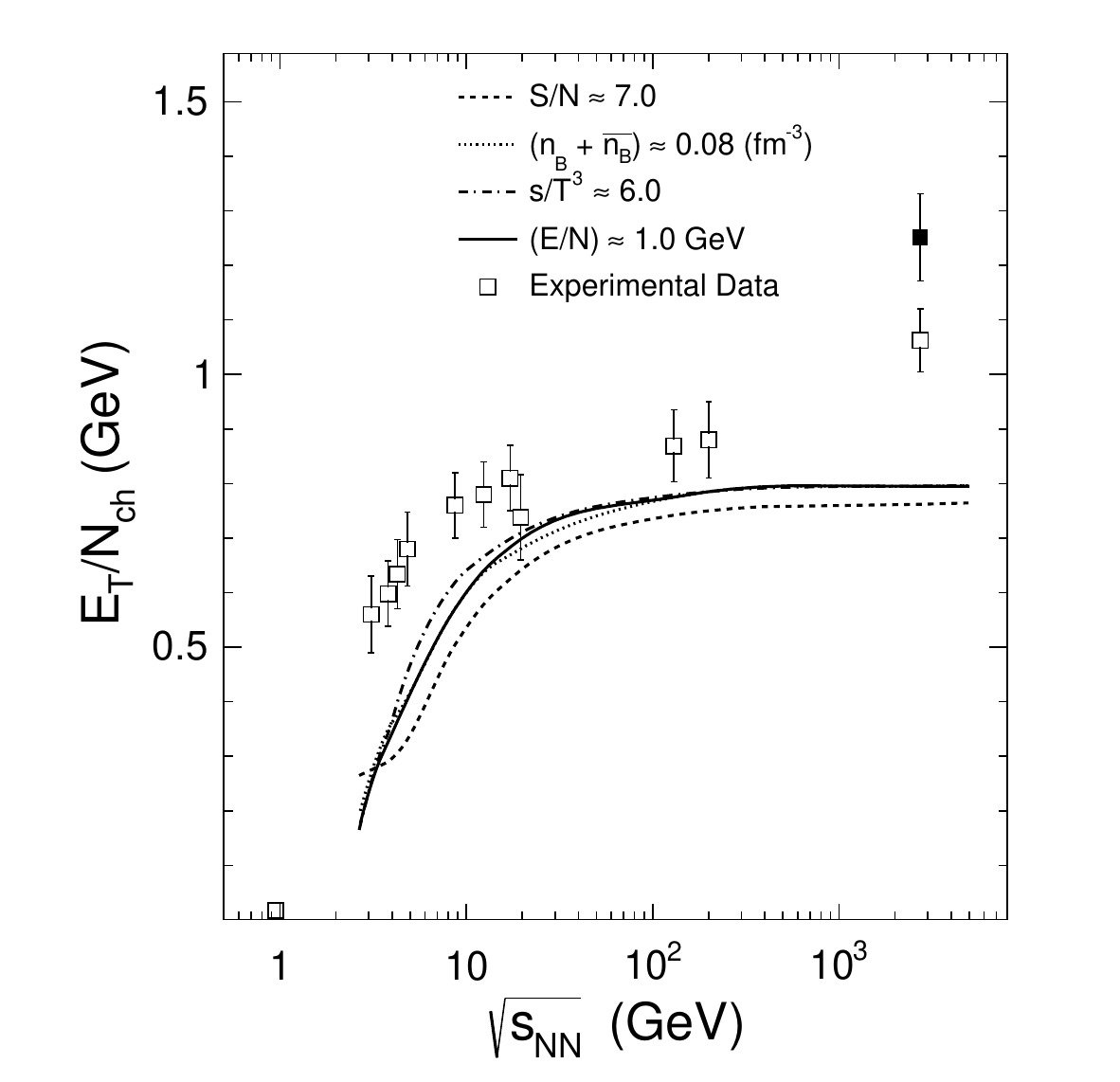}
\caption[]{The comparison between the experimental data on $E_{T}/N_{ch}$  and the expectations from various universal freeze-out criteria in the framework of an EV-SHGM.}
\label{et_nch_snn}
\end{figure}
In this section, we discuss the comparison of the experimental data on the ratio $E_{T}/N_{ch}$ with that calculated in EV-SHGM using various universal freeze-out criteria in heavy-ion collisions, such as the energy per hadrons (E/N)~\cite{Cleymans:1998fq,Cleymans:2005xv}, the sum of baryons and antibaryons ($n_{B}+n_{\bar{B}}$)~\cite{BraunMunzinger:2001mh}, the normalized entropy density, $s/T^3$~\cite{Tawfik:2004vv}, and the entropy per particle (S/N)~\cite{Tiwari:2012}. These observables are almost independent of $\sqrt{s_{NN}}$ except at lower energies. In figure~\ref{et_nch_snn}, we show the variations of $E_{T}/N_{ch}$  with $\sqrt{s_{NN}}$ from AGS to LHC energies. Here, the symbols are the experimental data while the lines are those calculated using various freeze-out criteria in our excluded-volume model. We find a similar behaviour between the experimental data and those calculated using various freeze-out criteria \cite{Cleymans:2008it} in our model upto top RHIC energy, while at LHC energies our calculations could not explain the experimental data. This points to further investigation(s) to understand the possible reason(s) for the deviation of LHC data from the universal freeze-out criteria.

\section{Summary and Conclusion}
\label{summary}
In summary, we have performed a calculation of global observables like transverse energy of hadrons, charged particle multiplicity and their ratios at mid-rapidity using an excluded-volume statistical-thermal model. We calculate the transverse momentum spectra of $\pi^-$ at various $\sqrt{s_{NN}}$ by using EV-SHGM. We get the radial flow velocity at various energies by comparing our calculations with the experimental data. We then estimate the transverse energy per unit rapidity and Bjorken energy density at different $\sqrt{s_{NN}}$. We study the centrality dependence of various hadron ratios using EV-SHGM and extract T and $\mu_{B}$ at RHIC and LHC energies. The estimated values of $T$ and $\mu_{B}$ are then used to study the centrality dependence of the ratio $E_{T}/N_{ch}$. Finally, we calculate the ratio $E_{T}/N_{ch}$ at various energies using EV-SHGM with and without flow. Further we study various freeze-out criteria in the framework of EV-SHGM using $E_{T}/N_{ch}$ as the observable. 

In conclusion, we have successfully described the $N_{part}$-dependence of various hadron ratios using the excluded-volume model. We observe that the inclusion of the collective flow in the model qualitatively explains the centrality data at the LHC, with some degree of deviations for higher centralities. While studying the energy dependence of $E_{T}/N_{ch}$, we observe that the EV-SHGM with collective flow does not explain the CMS data at the LHC, whereas the ALICE data at the same energy is well explained. Precision measurements of $E_{T}/N_{ch}$ at LHC energies are needed to see if mechanisms other than the collective flow play a role. The energy dependence of Bjorken energy density, pseudorapidity densities of charged particles and transverse energy and total charged particle multiplicity showing deviations from a logarithmic behaviour \cite{Mishra:2013dwa,Mishra:2014dta,Sarkisyan:2016dzo,Sarkisyan:2015gca,Abbas:2013bpa,Chatrchyan:2011pb,ATLAS:2011ag,Aamodt:2010pb} at LHC may indicate a different multiparticle production mechanism at the LHC, compared to lower collision energies. The observed increase in $E_{T}/N_{ch}$ from RHIC to LHC is attributed to an increase in $\langle p_{T} \rangle$ and the onset of higher collective flow at the LHC. Our comparison of the energy dependence of $E_{T}/N_{ch}$ with various universal freeze-out criteria reveals that further investigations are necessary in order to have a proper understanding of the LHC data and its connection with freeze-out. We give a prediction for the value of  $E_{T}/N_{ch}$ = 1.45 GeV with radial flow using EV-SHGM for $\sqrt{s_{NN}}$ = 5.02 TeV.

\section*{Acknowledgement}
 The authors acknowledge the financial supports from ALICE Project No. SR/MF/PS-01/2014-IITI(G) of Department of Science $\&$ Technology, Government of India.


\begin{thebibliography}{99}

\bibitem{Arsene:2004fa} 
  I.~Arsene {\it et al.} [BRAHMS Collaboration],
Nucl.\ Phys.\ A {\bf 757}, 1 (2005).
  
  
  \bibitem{Back:2004je} 
  B.~B.~Back {\it et al.},
  Nucl.\ Phys.\ A {\bf 757}, 28 (2005).
  
  
  \bibitem{Adams:2005dq} 
  J.~Adams {\it et al.} [STAR Collaboration],
  Nucl.\ Phys.\ A {\bf 757}, 102 (2005).
  
  
  
  \bibitem{Adcox:2004mh} 
  K.~Adcox {\it et al.} [PHENIX Collaboration],
  Nucl.\ Phys.\ A {\bf 757}, 184 (2005).
  
  
  \bibitem{Policastro:2001yc} 
  G.~Policastro, D.~T.~Son and A.~O.~Starinets,
  Phys.\ Rev.\ Lett.\  {\bf 87}, 081601 (2001).
  


\bibitem{Khachatryan:2016txc} 
  V.~Khachatryan {\it et al.} [CMS Collaboration],
  Phys.\ Lett. \ B {\bf 765}, 193 (2017).
  
  
  
  \bibitem{Abelev:2012rv} 
  B.~Abelev {\it et al.} [ALICE Collaboration],
  Phys.\ Rev.\ Lett.\  {\bf 109}, 072301 (2012).
  

\bibitem{Cleymans:2007uk} 
  J.~Cleymans, R.~Sahoo, D.~P.~Mahapatra, D.~K.~Srivastava and S.~Wheaton,
  Phys.\ Lett.\ B {\bf 660}, 172 (2008).
     
  

\bibitem{Mishra:2013dwa} 
R.~Sahoo, and A.~N.~Mishra,
  Int.\ J.\ Mod.\ Phys.\ E {\bf 23}, 1450024 (2014).
  
  
  
  \bibitem{Sahoo:2014aca} 
  R.~Sahoo, A.~N.~Mishra, N.~K.~Behera and B.~K.~Nandi,
  Adv.\ High Energy Phys.\  {\bf 2015}, 612390 (2015) and references therein.
  
  
  
  \bibitem{Bjorken:1982qr} 
  J.~D.~Bjorken,
  Phys.\ Rev.\ D {\bf 27}, 140 (1983).
  
 
 
 
  
    \bibitem{Prorok:2004af} 
  D.~Prorok,
  Eur.\ Phys.\ J.\ A {\bf 24}, 93 (2005).
  
  
  
  \bibitem{Prorok:2004wi} 
  D.~Prorok,
  Eur.\ Phys.\ J.\ A {\bf 26}, 277 (2005).
  
  
  
  \bibitem{Prorok:2006ve} 
  D.~Prorok,
  Phys.\ Rev.\ C {\bf 75}, 014903 (2007).
  

  \bibitem{Mishra:2008tc} M.~Mishra and C.~P.~Singh,
  Phys.\ Rev.\ C {\bf 78}, 024910 (2008).
  
  
\bibitem{Tiwari:2013wga} 
  S.~K.~Tiwari and C.~P.~Singh,
  Adv.\ High Energy Phys.\  {\bf 2013}, 805413 (2013). 
  
  
  
  \bibitem{Tiwari:2013pva} 
  S.~K.~Tiwari and C.~P.~Singh,
  J.\ Phys.\ Conf.\ Ser.\  {\bf 509}, 012097 (2014).
  
  
  \bibitem{Tiwari:2013} 
  S.~K.~Tiwari, P.~K.~Srivastava and C.~P.~Singh,
  J.\ Phys.\ G {\bf 40}, 045102 (2013).
  
  
  \bibitem{Adam:2016thv} 
  J.~Adam {\it et al.} [ALICE Collaboration],
  Phys.\ Rev.\ C {\bf 94}, 034903 (2016).
  
  
  \bibitem{Adams:2004cb} 
  J.~Adams {\it et al.} [STAR Collaboration],
  Phys.\ Rev.\ C {\bf 70}, 054907 (2004).


  
  \bibitem{Adler:2004zn} 
  S.~S.~Adler {\it et al.} [PHENIX Collaboration],
  Phys.\ Rev.\ C {\bf 71}, 034908 (2005)
  Erratum: [Phys.\ Rev.\ C {\bf 71}, 049901 (2005)].
  
  
  \bibitem{Cooper:1974mv} 
  F.~Cooper and G.~Frye,
  Phys.\ Rev.\ D {\bf 10}, 186 (1974).
  
  
  \bibitem{Yin:2017qhg} 
  X.~Yin, C.~M.~Ko, Y.~Sun and L.~Zhu,
  Phys.\ Rev.\ C {\bf 95}, 054913 (2017).
  
  \bibitem{qgp}
  K.~Yagi, T.~Hatsuda and Y.~Miake,
  Camb.\ Monogr.\ Part.\ Phys.\ Nucl.\ Phys.\ Cosmol.\  {\bf 23}, 1 (2005).
  
  
 
\bibitem{Kharzeev:2000ph} 
  D.~Kharzeev and M.~Nardi,
  Phys.\ Lett.\ B {\bf 507}, 121 (2001).
 
 
 \bibitem{Tiwari:2012} 
  S.~K.~Tiwari, P.~K.~Srivastava and C.~P.~Singh,
  Phys.\ Rev.\ C {\bf 85}, 014908 (2012).
  
  
  
  
  
  \bibitem{Adamczyk:2017iwn} 
  L.~Adamczyk {\it et al.} [STAR Collaboration],
  Phys.\ Rev.\ C {\bf 96}, 044904 (2017).
  
  
  \bibitem{Abelev:2007ra} 
  B.~I.~Abelev {\it et al.} [STAR Collaboration],
  Phys.\ Lett.\ B {\bf 655}, 104 (2007).
  
  \bibitem{Abelev:2006jr} 
  B.~I.~Abelev {\it et al.} [STAR Collaboration],
  Phys.\ Rev.\ Lett.\  {\bf 97}, 152301 (2006)
  
  
  \bibitem{Abelev:2014laa} 
  B.~B.~Abelev {\it et al.} [ALICE Collaboration],
  Phys.\ Lett.\ B {\bf 736}, 196 (2014).
  
  
%
%
%
%
  
  
  \bibitem{Chatterjee:2015fua} 
  S.~Chatterjee, S.~Das, L.~Kumar, D.~Mishra, B.~Mohanty, R.~Sahoo and N.~Sharma,
  Adv.\ High Energy Phys.\  {\bf 2015}, 349013 (2015) and references therein.
  
  
  \bibitem{Chatrchyan:2012mb} 
  S.~Chatrchyan {\it et al.} [CMS Collaboration],
  Phys.\ Rev.\ Lett.\  {\bf 109}, 152303 (2012).
  
  
  \bibitem{Eskola:1999fc} 
  K.~J.~Eskola, K.~Kajantie, P.~V.~Ruuskanen and K.~Tuominen,
  Nucl.\ Phys.\ B {\bf 570}, 379 (2000).


  
  
  \bibitem{Arsene:2005mr} 
  I.~Arsene {\it et al.} [BRAHMS Collaboration],
  Phys.\ Rev.\ C {\bf 72}, 014908 (2005).
  
  
  \bibitem{Cleymans:2004pp} 
  J.~Cleymans, B.~Kampfer, M.~Kaneta, S.~Wheaton and N.~Xu,
  Phys.\ Rev.\ C {\bf 71}, 054901 (2005).
  
  
  \bibitem{Abelev:2013vea} 
  B.~Abelev {\it et al.} [ALICE Collaboration],
  Phys.\ Rev.\ C {\bf 88}, 044910 (2013).
  
  
   
  \bibitem{Heinz:2007in} 
  U.~W.~Heinz and G.~Kestin,
  Eur.\ Phys.\ J.\ ST {\bf 155}, 75 (2008).
  



   
  \bibitem{Adams:2003xp} 
  J.~Adams {\it et al.} [STAR Collaboration],
  Phys.\ Rev.\ Lett.\  {\bf 92}, 112301 (2004).
  
 
  
%
%
  
  
   
 \bibitem{Cleymans:1998fq} 
  J.~Cleymans and K.~Redlich,
  Phys.\ Rev.\ Lett.\  {\bf 81}, 5284 (1998).
  
  
  \bibitem{Cleymans:2005xv} 
  J.~Cleymans, H.~Oeschler, K.~Redlich and S.~Wheaton,
  Phys.\ Rev.\ C {\bf 73}, 034905 (2006).
  
  
  \bibitem{BraunMunzinger:2001mh} 
  P.~Braun-Munzinger and J.~Stachel,
  J.\ Phys.\ G {\bf 28}, 1971 (2002).
  
  
  \bibitem{Tawfik:2004vv} 
  A.~Tawfik,
  J.\ Phys.\ G {\bf 31}, S1105 (2005).
  
  
  
    \bibitem{Cleymans:2008it} 
  J.~Cleymans, R.~Sahoo, D.~P.~Mahapatra, D.~K.~Srivastava and S.~Wheaton,
  J.\ Phys.\ G {\bf 35}, 104147 (2008).
  
  \bibitem{Mishra:2014dta} 
  A.~N.~Mishra, R.~Sahoo, E.~K.~G.~Sarkisyan and A.~S.~Sakharov,
  Eur.\ Phys.\ J.\ C {\bf 74}, 3147 (2014)
  Erratum: [Eur.\ Phys.\ J.\ C {\bf 75}, 70 (2015)].
  
 
  \bibitem{Sarkisyan:2016dzo} 
  E.~K.~G.~Sarkisyan, A.~N.~Mishra, R.~Sahoo and A.~S.~Sakharov,
  Phys.\ Rev.\ D {\bf 94}, 011501 (2016).
  
  \bibitem{Sarkisyan:2015gca} 
  E.~K.~G.~Sarkisyan, A.~N.~Mishra, R.~Sahoo and A.~S.~Sakharov,
  Phys.\ Rev.\ D {\bf 93}, 054046 (2016);
  Addendum: [Phys.\ Rev.\ D {\bf 93}, no. 7, 079904 (2016)].




\bibitem{Abbas:2013bpa} 
  E.~Abbas {\it et al.} [ALICE Collaboration],
  Phys.\ Lett.\ B {\bf 726}, 610 (2013).
  
  
  
  
  \bibitem{Chatrchyan:2011pb} 
  S.~Chatrchyan {\it et al.} [CMS Collaboration],
  JHEP {\bf 1108}, 141 (2011).
  
  
  \bibitem{ATLAS:2011ag} 
  G.~Aad {\it et al.} [ATLAS Collaboration],
  Phys.\ Lett.\ B {\bf 710}, 363 (2012).
  
  
  \bibitem{Aamodt:2010pb} 
  K.~Aamodt {\it et al.} [ALICE Collaboration],
  Phys.\ Rev.\ Lett.\  {\bf 105}, 252301 (2010).





\end{thebibliography}
\end{document}